\documentclass[12pt]{article}
\usepackage{amssymb}

\newcommand{\newsection}{ \setcounter{equation}{0} \section}
\newcommand*\eq{\begin{equation}}
\newcommand*\en{\end{equation}}
\newcommand*\eqa{\begin{eqnarray}}
\newcommand*\ena{\end{eqnarray}}
\def\nn{\nonumber \\}
\def\al{\alpha}
\def\ve{\varepsilon}
\def\C{{\mathbb C}}
\def\Ax{{\mathcal{A}_x}}
\def\dc{\stackrel{\diamond}{,}}
\def\ds{\stackrel{*}{,}}
\def\x{\hat x}
\def\y{\hat y}
\def\xx{{\hat X}}
\def\pat{\partial}
\def\c#1{{\cal #1}}
\def\ad{\mbox{ad}\,}

  \newcommand*\ti[5]{{\em #5}, {#1} {\bf #2}, #3 (#4)}
\providecommand{\href}[2]{#2}
\newcommand*\xxx[1]{\href{http://xxx.lanl.gov/abs/#1}{{#1}}}

\newcommand*\jhep{JHEP}
\newcommand*\np{Nucl. Phys.}

\newcommand*\pr{Phys. Rev.}

\begin{document}

\begin{titlepage}
\rightline{LMU-TPW 2000-05}
\rightline{MPI-PhT/2000-04}
\rightline{hep-th/0001203}

\vspace{4em}
\begin{center}

 {\Large{\bf Gauge Theory on Noncommutative Spaces}}

 \vskip 3em

{{\bf
J.\ Madore${}^{1,2}$, S.\ Schraml${}^{2,3}$, P.\ Schupp${}^{3}$, 
J.\ Wess${}^{2,3}$ }}

 \vskip 1em

${}^1$Laboratoire de Physique Th\'eorique\\
        Universit\'e de Paris-Sud, B\^atiment 211, F-91405 Orsay\\[1em]
${}^{2}$Max-Planck-Institut f\"ur Physik\\
        F\"ohringer Ring 6, D-80805 M\"unchen\\[1em]
${}^{3}$Sektion Physik, 
Universit\"at M\"unchen\\
Theresienstr.\ 37,
D-80333 M\"unchen

 \end{center}

 \vspace{2em}

\begin{abstract} 
We introduce a formulation of gauge theory on noncommutative spaces 
based on the concept of covariant coordinates. Some important examples 
are discussed in detail. A Seiberg-Witten map is established in all cases.
\end{abstract}

\vfill
%\noindent \hrule
%\hbox{{\small{\it e-mail: }}{\small\quad 
% \quad schupp@theorie.physik.uni-muenchen.de}}
\end{titlepage}\vskip.2cm

\newpage

\setcounter{page}{1}

\newsection{Introduction}

We introduce a natural method to formulate a gauge theory on more or 
less arbitrary noncommutative spaces. The starting point is the       
observation that multiplication of a (covariant) field by a coordinate 
can in general not be a covariant operation in noncommutative geometry, 
because the coordinates will not commute with the gauge 
transformations. 
The idea is to make the coordinates covariant by adding a gauge 
potential to them.  This is analogous to the case in usual gauge 
theory; one adds gauge potentials to the partial derivatives to obtain 
covariant derivatives. 
One can consider a covariant coordinate as a position-space 
analogue of the usual covariant momentum of gauge theory.

In the following we prefer not to present the general case of an 
arbitrary associative algebra of noncommuting variables; we consider  
rather three important examples in which the commutator of two 
coordinates is respectively constant, linear and quadratic in the 
coordinates.  We employ Weyl's quantization proceedure to associate 
with an algebra of noncommuting coordinates an algebra of functions of 
commuting variables with deformed product.  One of our examples gives 
the same kind of noncommutative gauge theory that has 
surfaced in string theory recently~\cite{SW}.

\newsection{Covariant coordinates}\label{sec:Covar-coord}

The associative algebraic structure $\Ax$ which defines a noncommutative space
can be defined in terms of a set of generators $\x^i$ and relations
$\c{R}$.  Instead of considering a general expression for the
relations we shall discuss rather some important explicit cases. These
are of the form of 
a canonical structure
\eq
[\x^i,\x^j] = i \theta^{ij} , \qquad \theta^{ij} \in \C,          \label{can}
\en
a Lie-algebra structure
\eq
[\x^i,\x^j] = i C^{ij}{}_k \x^k , \qquad C^{ij}{}_k \in \C,           \label{lie}
\en
and a quantum space structure \cite{schladming,calculus,cov_diff,SWZ}
\eq
\x^i \x^j = q^{-1}\hat R^{ij}{}_{kl} \x^k \x^l , \qquad 
\hat{R}^{ij}{}_{kl} \in \C.                                   \label{qua}
\en
In all these cases the index $i$ takes values from $1$ to $N$. We
shall suppose that $\Ax$ has a unit element. For the quantum space 
structure a simple version is the Manin plane, 
with $N=2$:
\eq
\x \y = q \y \x , \qquad q \in \C.                         \label{man}
\en
We shall refer to the generators $\x^i$ of the algebra as `coordinates' 
and we shall consider $\Ax$ to be the algebra of formal power series 
in the coordinates modulo the relations
\eq
\Ax \equiv \C \left[[\x^1, \ldots, \x^N]\right] / \c{R}.           \label{rel}
\en
For a physicist this means that one is free to use the relations
(\ref{can}), (\ref{lie}) or (\ref{qua}), (\ref{man}) to reorder the 
elements of an arbitrary power series.

We consider fields as elements of the algebra  $\Ax$:
\eq
\psi(\x) = \psi(\x^1, \ldots, \x^N) \in \Ax.                     \label{field}
\en
We shall introduce the notion of an infinitesimal gauge transformation 
$\delta\psi$ of the field $\psi$ and suppose that under an
infinitesimal gauge transformation $\al(\x)$ it can be written in the
form 
\eq
\delta \psi(\x) = i \al(\x) \psi(\x) ; \qquad \al(\x), \; \psi(\x) \in \Ax.
\label{transform}
\en
This we call a covariant transformation law of a field. 
It follows then of course that $\delta\psi \in \Ax$. Since $\al(\x)$ is
an element of $\Ax$ it is the equivalent of an abelian gauge
transformation. If $\al(\x)$ belonged to an algebra $M_n(\Ax)$ of
matrices with elements in $\Ax$ then it would be the equivalent of a
nonabelian gauge transformation.  

An essential concept is that the coordinates are 
invariant under the action of a gauge transformation:
$$
\delta \x^i = 0.
$$
Multiplication of a field on the left by a coordinate is then
not a covariant operation in the noncommutative case. That is
\eq
\delta (\x^i \psi) = i \x^i \al(\x) \psi
\en
and in general the right-hand side is not equal to $i\al(\x)\x^i\psi$.

Following the ideas of ordinary gauge theory we introduce covariant 
coordinates $\xx^i$ such that
\eq
\delta (\xx^i \psi) = i \al \xx^i \psi ,                  \label{cov}
\en
i.e., $\delta(\xx^i) = i[\alpha , \xx^i]$.
To find the relation between $\xx^i$ and $\x^i$ we make
an Ansatz of the form
\eq
\xx^i = \x^i + A^i(\x) , \qquad A^i(\x) \in \Ax.                 \label{ans}
\en
This is quite analogous to the expression of a covariant derivative as
the sum of an ordinary derivative plus a gauge potential.%
\footnote{Closely related to the coordinate $\x^i$ is the inner derivation 
$\ad \x^i$ of $\Ax$ and in this context a general consistency relation 
for $\x^i$ has been written~\cite{DimMad96} which also covers the
relations (\ref{can}), 
(\ref{lie}) and (\ref{qua}).}

We derive the transformation properties of $A^i$ from the requirement 
(\ref{cov}):
\eq
\delta A^i = i [\al, A^i] - i [\x^i,\al] .                  \label{gauge}
\en
The right hand side can be evaluated using one of the relations
(\ref{can}), (\ref{lie}) or (\ref{qua}). It is easy to see that
a tensor $T^{ij}$ can be defined in each case as respectively
\eq
 T^{ij} = [\xx^i , \xx^j] - i \theta^{ij}                   \label{tcan}
\en
in the canonical case,
\eq
T^{ij} = [\xx^i , \xx^j ] - i C^{ij}{}_k \xx^k           \label{tlie}
\en
for the Lie-structure
and
\eq
T^{ij} = \xx^i \xx^j - 
q^{-1} \hat{R}^{ij}{}_{kl} \xx^k \xx^l                      \label{tqua}
\en
for the quantum space.%
\footnote{The second expression (\ref{tlie})
has a direct interpretation as the field strength of an electromagnetic
potential over a geometry with $\Ax = M_n$, the algebra of 
$n \times n$ matrices~\cite{DubKerMad89a}. 
It is of interest to note that in the case where
the $\x^i$ are used to construct inner derivations then the analog of
$T^{ij}$ must vanish~\cite{DimMad96,Mad99c}.}

We verify directly that the objects $T^{ij}$ are covariant tensors. In the canonical case
we find
\eqa
T^{ij} & = & [A^i, \x^j] + [\x^i,A^j] + [A^i,A^j] ,\nonumber \\
\delta T^{ij} & = & [\delta A^i,\x^j] + [\x^i, \delta A^j]
+[\delta A^i, A^j] + [A^i, \delta A^j] .  \label{cantensor}
\ena
We insert $\delta A^i$ from (\ref{gauge}), use the Jacobi identity and obtain
\eq
\delta T^{ij} = i [\al, T^{ij}] . \label{tensor}
\en
Exactly the same procedure leads to the result for the Lie structure:
\eqa
T^{ij} & = & [\x^i , A^j] + [A^i, \x^j] + [A^i, A^j] - i C^{ij}{}_k A^k , \nonumber\\
\delta T^{ij} & = & i [\al, T^{ij}] .
\ena
In the case of the quantum space we find
\eq
T^{ij} = P^{ij}{}_{kl}(A^k \x^l + \x^k A^l + A^k A^l)
\en
where we have introduced $P$ defined as
\eq
P^{ij}{}_{kl} = \delta^i_k\delta^j_l 
                - q^{-1} \hat R^{ij}{}_{kl}.
\en
%(The fact that $P/2$ is a projector is equivalent to the
%condition
%$q^{-2}\hat R^2 = I$, 
%which we must impose on the braid matrix $\hat R$.)
We again insert $\delta A^i$ from (\ref{gauge}) to compute $\delta T^{ij}$.
We obtain
\eqa
\delta T^{ij} & = & i P^{ij}{}_{kl} \left\{ [\al, A^k] \x^l + [\al, \x^k]\x^l
+\x^k[\al, A^l] + \x^k[\al,\x^l] \right.\nonumber \\[4pt]
&& \left.+[\al,A^k]A^l + [\al,\x^k]A^l + A^k[\al,A^l] + A^k[\al,\x^l]\right\} .
\ena
With relation (\ref{qua}) this becomes
\eq
\delta T^{ij} = i [\al, T^{ij}] .
\en

\newsection{Weyl Quantization}

In the framework of canonical quantization Hermann Weyl \cite{Weyl} gave a
prescription how to associate an operator with a classical function of
the canonical variables.  This prescription can also be used to
associate an element of $\Ax$ with a function $f$ of classical
variables $x^1, \ldots x^n$ \cite{Wigner}. We use $\x$ for elements of
$\Ax$ and $x$ for the associated classical commuting variables.

Using the Fourier transform
\eq
\tilde{f}(k) = 
\frac{1}{(2\pi)^{\frac{n}{2}}} \int d^nx\, e^{-ik_j x^j} f(x)   \label{fourier}
\en
 of the function $f(x^1, \ldots x^n)$ we define an operator
\eq
W(f)=\frac{1}{(2\pi)^{\frac{n}{2}}}
\int d^nk\, e^{ik_j \x^j}\tilde{f}(k).                  \label{weylqua}
\en
This is a unique prescription, the operator $\x$ replaces the
variables $x$ in $f$ in the most symmetric way. If the operators $\x$ 
have hermiticity properties $W(f)$ will inherit these properties
for real $f$. At present we are interested in the algebraic properties
only.

Operators obtained by (\ref{weylqua}) can be multiplied to yield new
operators.   The question arises whether or not these new operators can 
be associated also with classical functions. 
If such a function exists we call it $f\diamond g$ (`f diamond g'):
\eq
  W(f)W(g)=W(f\diamond g).                             \label{weylprod}
\en
We can write (\ref{weylprod}) more explicitly as
\eq
\label{weylprodexp}
W(f)W(g)=\frac{1}{(2\pi)^{n}}\int d^nk d^np\, 
e^{ik_i\x^i}e^{ip_j\x^j}\tilde{f}(k)\tilde{g}(p).
\en
If the product of the two exponentials can be calculated by the 
Baker-Campbell-Hausdorff formula to give an exponential of a linear combination of
the $\x^i$ the function $f\diamond g$ will exist.

This is the case for the canonical structure:
\eq
e^{ik_i\x^i} e^{ip_j \x^j} = 
e^{i(k_j+p_j)\x^j-\frac{i}{2}k_ip_j\theta^{ij}}.        \label{eprod}
\en
A comparison with (\ref{weylqua}) shows that $(f\diamond g)(x)$ can be computed from  
(\ref{weylprodexp}) and (\ref{eprod}) by replacing $\x$ by $x$.
\eqa
f\diamond g = f\ast g &=& 
\frac{1}{(2\pi)^{n}}\int d^nk d^np\,e^{i(k_j+p_j)x^j -
\frac{i}{2} k_i\theta^{ij}p_j}\tilde{f}(k)\tilde{g}(p)\nonumber\\[4pt]
&=& \left. e^{\frac{i}{2}\frac{\pat}{\pat x^i}           \label{moyal}
  \theta^{ij}\frac{\pat}{\pat y^j}}f(x)g(y)\right|_{y\to x}
\ena
We obtain the Moyal-Weyl $\ast$-product \cite{Moyal}.

A similar $\ast$-product is obtained for the Lie structure:
\eq
e^{ik_i\x^i} e^{i p_j\x^j} = e^{iP_i(k,p)\x^i}
\en
where $P_i(k,p)$ are the parameters of a group element obtained by 
multiplying two group elements, one parametrized by $k$ and the other 
by $p$. {From} the Baker-Campbell-Hausdorff formula we know that
\eq
P_i(k,p)=k_i+p_i+\frac{1}{2}g_i(k,p)
\en
where $g_i$ contains the information about the noncommutative structure of
the group.
Again we find the star product after a Fourier transformation 
\eqa
f\diamond g = f\ast g & = & 
\frac{1}{(2\pi)^{n}}\int d^nk d^np\,
e^{iP_i(k,p) x^i}\tilde{f}(k)\tilde{g}(p)\nonumber\\[4pt]
&=& \left. e^{\frac{i}{2}x^i\,
  g_i(i\frac{\pat}{\pat y},i\frac{\pat}{\pat z})}
  f(y)g(z)\right|_{y\to x \atop z\to x}.                  \label{starlie}
\ena

A more complicated situation arises for the quantum plane structure.
The Baker-Campbell-Hausdorff formula cannot be used explicity. The Weyl
quantization (\ref{weylqua}) does not seem to be the most natural
one. At the moment we are only interested in the algebraic structure
of the theory. In this context any unique way of associating an
operator with a function of the classical variables would do. For the
quantum plane this could be a normal ordering. We treat the case of 
the Manin plane (\ref{man}) explicitly.

With any monomial in $x$ $y$ we associate the normal ordered product 
of the operators $\x$, $\y$ where all the $\x$ operators
are placed to the left and all the $\y$ operators to the right:
\eq
W(f(x,y))=\; :f(\x,\y):
\en
The dots indicate the above normal ordering.
Eqn (\ref{weylprod}) now has to be written in the form:
\eq
:f(\x,\y):\;:g(\x,\y):\;=\; :f\diamond g(\x,\y):
\en
Let us first compute this for monomials:
\eqa
\x^{n_1} \y^{m_1} \x^{n_2} \y^{m_2} &=& 
q^{-m_1n_2}\x^{n_1+n_2} \y^{m_1+m_2}\\[4pt]
:\x^{n_1} \y^{m_1}:\; :\x^{n_2} \y^{m_2}:\; &=& 
q^{-m_1n_2}\;:\x^{n_1+n_2} \y^{m_1+m_2}:\nonumber\\[4pt]
&=& W\left(\left. q^{-x'\frac{\pat}{\pat x'}y
  \frac{\pat}{\pat y}} x^{n_1}y^{m_1}x'^{n_2}y'^{m_2}
  \right|_{x'\to x \atop y'\to y}\right)\nonumber
\ena
This is easily generalized to arbitrary power series in $x$ and $y$
\eq
\label{qua_diamond}
f\diamond g = \left. q^{-x'\frac{\pat}{\pat x'}y
  \frac{\pat}{\pat y}}f(x,y)g(x',y')\right|_{x'\to x \atop y'\to y}
\en
and we have obtained a diamond product for the Manin plane.
Instead of the $\x \y$ ordering we could have used the 
$\y \x$ ordering or, more reasonably, the totally symmetric
product of the $\x \y$ operators. For monomials of fixed
degree the $\x \y$ ordered and the $\y \x$ ordered as
well as the symmetrically ordered products form a basis. Thus the
diamond product exists in all the cases and it is only a combinatorial
problem to compute it explicitly.

The Weyl quantization allows the representation of an element of $\Ax$ by a
classical function of $x$. For a constant $c$ and for $\x^i \in \Ax$ this is trivial:
\eq
c   \rightarrow  c , \qquad
\x^i \rightarrow x^i .
\en
The formula (\ref{weylprod}) can be used to generalize this to any element of $\Ax$. As
an example we take the bilinear elements of $\Ax$.
\eq
\x^i \x^j = W(x^i) W(x^j) = W(x^i \diamond x^j), 
\qquad \x^i \x^j \rightarrow x^i \diamond x^j .
\en
In particular
\eq
W(x^i x^j) = \frac{1}{2}(\x^i \x^j + \x^j \x^i)
\en
for the canonical structure and the Lie structure. For the quantum space
structure we have
\eq
W(x^i x^j) =\; :\x^i \x^j: 
\en
The elements of $\Ax$ can be represented by functions $f(x)$, the multiplication
of the elements by the star product of the functions. This product is
associative. Let us now represent a field by a classical function $\psi(x)$. The
gauge transformation (\ref{transform}) is represented by $\al(x)$:
\eq
\delta_\al \psi(x) = i \al(x) \diamond \psi(x) .
\en
We immediately conclude
\eqa
(\delta_\al \delta_\beta - \delta_\beta \delta_\al) \psi(x) & = &
i \beta(x) \diamond \left(\al(x) \diamond \psi(x)\right) -
i \al(x) \diamond \left(\beta(x) \diamond \psi(x)\right) \nn
& = & i\left(\beta \diamond \al - \al \diamond \beta\right) \diamond \psi .
\ena
The transformation law of $A^i(x)$, representing the element $A^i \in \Ax$ is:
\eq
\label{gaugedia}
\delta A^i = i [\al \dc A^i ] - i [x^i \dc \al]
\en
and for the tensors $T^{ij}(x)$:
\eq
\delta T^{ij} = i[\al \dc T^{ij}].
\en
Where $T^{ij}$ is defined as in (\ref{tcan}), (\ref{tlie}), (\ref{tqua}), 
but with elements of $\Ax$ and algebraic multiplication 
replaced by the corresponding functions and diamond product:
\eqa
 T^{ij} &=& [A^i\dc x^j] + [x^i\dc A^j] + [A^i\dc A^j]  \nonumber\\
 T^{ij} &=& [x^i \dc A^j] + [A^i\dc x^j] + [A^i\dc A^j] - i C^{ij}{}_k A^k   \\
 T^{ij} &=& P^{ij}{}_{kl}(A^k\diamond x^l + x^k \diamond A^l + A^k\diamond A^l).               \nonumber
\ena

\newsection{Noncommutative gauge theories}\label{sec:Nonc-gauge-theor}

\subsection{Canonical structure}

We would now like to give explicit formulas for the gauge transformation
and tensor in the canonical case and will explain the relation to 
the conventions of noncommutative Yang-Mills theory as presented in~\cite{SW}. 
The commutator $[\x^i, . ]$ in the transformation of a gauge potential 
(\ref{gauge}),
$$\delta A^i = -i[\x^i, \al] + i [\al, A^i],$$
acts as a derivation on elements of $\Ax$.
Due to the special form of the commutation relations
(\ref{can}) with the
\emph{constant} $\theta^{ij}$, this commutator can in fact be written
as a derivative on elements $f \in \Ax$:
\eq
[\x^i, f] = i \theta^{ij} \pat_j f . \label{deriv}
\en
The derivative $\pat_j$ is defined as a derivation on $\Ax$, i.e.,
$\pat_j f g  = (\pat_j f)g + f (\pat_j g)$ and on the
coordinates as: $\pat_j \x^i \equiv \delta_j^i$. 
The right-hand side of (\ref{deriv}) is a derivation because $\theta$ 
is constant and thus commutes with everything. 
We find that in the canonical case
the gauge transformation can be written
\eq
\delta A^i = \theta^{ij} \pat_j \alpha + i [\alpha, A^i] .
\en
The gauge potential $\hat A$ of noncommutative Yang-Mills is 
introduced by the identification
\eq
A^i \equiv \theta^{ij} \hat A_j . \label{hata}
\en
We must here assume that the matrix $\theta$ is non-degenerate. 
We find the following transformation law for the gauge field $\hat A_j$:
\eq
\delta \hat A_j = \pat_j \alpha + i[\alpha,\hat A_j] .
\en
It has exactly the same form as the transformation law for a 
non-abelian gauge potential in commutative geometry, except that in 
general the meaning of the commutator is different. 
An explicit expression for the
tensor $T$ in the canonical case (\ref{cantensor}) is found likewise, 
\eq
T^{ij} = i \theta^{ik} \pat_k A^j - i \theta^{jl} \pat_l A^i + [A^i,A^j] .
\en
Up to a factor $i$, 
the relation to the field strength $\hat F$ of noncommutative Yang-Mills is
again simply obtained by using $\theta$ to raise indices:
\eq
T^{ij} = i \theta^{ik} \theta^{jl} \hat F_{kl} . \label{hatf}
\en
Assuming again non-degeneracy of $\theta$, we find
\eq
\hat F_{kl} = \pat_k \hat A_l - \pat_l \hat A_k 
- i [\hat A_k , \hat A_l] .
\en
According to our conventions we are to consider this as the field 
strength of an abelian gauge potential in a noncommutative geometry, 
but except for the definition of the bracket it has again the same 
form as a non-abelian gauge field-strength in commutative geometry.
Since $\theta^{ij} \in \C$, $\hat F$ is a tensor:
\eq
\delta \hat F_{kl} = i[\alpha, \hat F_{kl}].
\en
These formulae become clearer and the relation to noncommutative Yang-Mills 
theory is even more direct, if we represent the elements of $\Ax$ by 
functions of the classical variables $x^i$ and use the Moyal-Weyl star product 
(\ref{moyal}). In particular equation (\ref{deriv}) becomes
\eq
x^i * f - f * x^i = i \theta^{ij} \pat_j f , 
\en
where $f(x)$ is now a function and 
$\pat_j f = \pat f /\pat x^j$ is the ordinary
derivative. 
This follows directly from 
the Moyal-Weyl product (\ref{moyal}).  The identifications 
(\ref{hata},\ref{hatf}) have the same form as before.  The relevant 
equations written in terms of the star product become
\eqa
\delta A^i & = & \theta^{ij} \pat_j \alpha + i\alpha * A^i - i A^i * \alpha ,\\
T^{ij} & = & i \theta^{ik} \pat_k A^j - i \theta^{jl} \pat_l A^i 
+ A^i * A^j -  A^j * A^i ,\\
\delta T^{ij} & = & i\alpha * T^{ij} - i T^{ij} * \alpha ,\\
\delta \hat A_j & = & \pat_j \alpha + i \alpha * \hat A_j
-i\hat A_j * \alpha ,\\
\hat F_{kl} &  = & \pat_k \hat A_l - \pat_l \hat A_k 
- i \hat A_k * \hat A_l  + i \hat A_l * \hat A_k ,\\
\delta \hat F_{kl} & = & i\alpha * \hat F_{kl} - i\hat F_{kl} * \alpha 
\ena
and
\eq
\delta_\al \delta_\beta - \delta_\beta \delta_\al 
 =  \delta_{(\beta * \al - \al *\beta)} .
\en
All this clearly generalizes to $A^i$, $\alpha$, $\hat A_j$ and $\hat F_{kl}$
that are (hermitean) $n \times n$ matrices. We will have to say more about that
later. It is interesting to note the form of the covariant 
coordinates written in terms of $\hat A$:
\eq
\xx^i = \x^i + \theta^{ij}\hat A_j  .
\en
This expression has appeared in string theory contexts related
to noncommutative Yang-Mills theory mainly as a coordinate transformation
\cite{Cornalba, Ishibashi, JS}.

\paragraph{Remark:} Ordinary gauge theory can be understood as a special
case of gauge theory on the noncommutative canonical structure as follows:
Consider coordinates $\{\hat q^j,\hat p_i\}$ with canonical commutation relations
$[\hat q^j, \hat p_i] = i \delta_i^j$ and restrict the allowed choices of
infinitesimal gauge transformations $\alpha$ to depend only on the 
$\hat q^i$, i.e., only
on half the original coordinates. Multiplying a field $\psi$ by
a coordinate is now a noncovariant concept only for half the coordinates,
namly for the `momenta' $\hat p_i$. The gauge field $A$ will thus depend only 
on the $\hat q^i$, as will the tensor $T$. It is not hard to see that the
relations of noncommutative gauge theory reduce in this case to those of
ordinary gauge theory. The algebra of the $\hat q^j$ and 
$\hat p_i$ can of course be realized as ordinary commutative 
coordinates $q^j$ and derivatives
$-i \pat_i$.

\subsection{Lie structure}

The relations of noncommutative gauge theory on a Lie structure (\ref{lie})
written in the language of star products are
\eqa
\delta A^i & = &  - i [x^i \ds \alpha]  + i[\alpha \ds A^i] , \label{lie1}\\
T^{ij} & = &  [x^i\ds A^j] + [A^i\ds x^j] + [A^i \ds A^j] - iC^{ij}{}_k A^k ,
\label{lie2}\\
\delta T^{ij} & = & i\alpha * T^{ij} - i T^{ij} * \alpha ,
\ena
where $A^i$ and $\alpha$ are functions of the (commutative) coordinates $x^i$
and the $\ast$-product is given in (\ref{starlie}).
As in the canonical case, $[x^i \ds f(x)]$ can be written in terms of a derivative
of $f$
\eq
[x^i \ds f(x)] = i C^{ij}{}_k  x^k \frac{\pat f}{\pat x^j} ,
\en
but the proof is not so obvious, because the left-hand side is a derivation 
of the noncommutative $*$-product while
the right-hand side is a derivation with respect to the commutative pointwise
product of functions. However, these two notions can be reconciled thanks
to the symmetrization inherent in the Weyl quantization proceedure.
Equations (\ref{lie1}) and (\ref{lie2}) can thus also be written as
\eqa
\delta A^i & = & C^{ij}{}_k x^k \pat_j \al + i \alpha * A^i - i A^i * \alpha ,
\\
T^{ij} & = & i C^{il}{}_k x^k \pat_l A^j
- i C^{jl}{}_k x^k \pat_l A^i + [A^i \ds A^j] - iC^{ij}{}_k A^k .
\ena

\newsection{Nonabelian gauge transformations}

In this case the parameter $\al(\x)$ in (\ref{transform}) and the gauge field $A$ in
(\ref{ans}) will be matrix valued:\footnote{For notational simplicity we are
suppresing the index $i$ on $A^i$.} $\al =  \al_r T^r$ and $A = A_r T^r$, where
$\al_r, A_r \in \Ax$ and the $T^r$ form a suitable basis of matrices.
It is not clear what conditions we can consistently impose on these matrices and in
particular in which sense they can be Lie-algebra valued; we can, however, always assume
that $\alpha$ and $A$ are in the enveloping algebra of a 
Lie algebra. Let us consider the commutator (\ref{gauge}). It can be written
as a sum of commutators and anticommutators of the matrices $T^i$:
\eq
[\alpha, A] = \frac{1}{2}(\alpha_r A_s + A_s \alpha_r)[T^r,T^s]
+ \frac{1}{2}(\alpha_r A_s - A_s \alpha_r)\{T^r,T^s\} . \label{comantcom}
\en
In the commutative case the second term is zero and it is clear that one can
choose $T_r$ from any matrix representation of a Lie algebra. Here, however,
$\al_r$ and $A_s$ do not commute. As we shall see it is nevertheless possible
to consistently impose hermiticity, while it is e.g.\ not consistent to impose
tracelessness.

Let us now assume that the relations 
(\ref{can}), (\ref{lie}), (\ref{qua}) or (\ref{man})
allow a conjugation:
\eq
(\x^i)^* = \x^i
\en
This will be the case for real $\theta^{ij}$, real $C^{ij}{}_k$ and, in (\ref{man}), $q$ a root of unity. Then it makes sense to speak about ``real'' functions
\eq
f^*(\x) = f(\x) ,
\en
and in this case $\alpha$ could be hermitean:
\eq
\alpha(\x) = \alpha_l(\x) T^l = \alpha^*(\x), \qquad
(\alpha_l(\x))^* = \alpha_l(\x), \qquad T_l^\dagger = T_l .
\en
The commutation of those hermitean objects will be antihermitean:
\eq
\left([\al(x), \beta(y)]\right)^* = -[\alpha(x), \beta(y)] .
\en
We conclude that with $\alpha$, $A$ and $\x$ hermitean, $\delta A$ in 
(\ref{gauge}) will be hermitean again. If the matrices $T_l$ form a basis for all 
hermitean matrices of
a certain dimension, then the commutators and anticommutators 
in (\ref{comantcom}) will also close into
these matrices.

\newsection{Seiberg-Witten map}

Seiberg and Witen were able to establish a connectoin of noncommutative 
Yang-Mills theory to ordinary Yang-Mills theory. We show that this can 
be done for all three examples we have considered. 

The ordinary gauge potential we shall call $a_i$ and the infinitesimal 
gauge parameter $\ve$. The transformation law of the gauge potential $a_i$ is
\begin{equation}
  \label{ordg}
  \delta_{\ve} a_i=\pat_i\ve+i[\ve,a_i].
\end{equation}
This has to be compared with the gauge transformation (\ref{gaugedia})
\begin{equation}
  \label{dia_delta_A}
  \delta A^i = i [\al \dc A^i ] - i [x^i \dc \al].
\end{equation}

The diamond product can be written in a formal way analogous to 
deformation quantization \cite{BFFLS,Kontsevich} 
\begin{equation}
  \label{def_qua}
  f\diamond g=fg+\sum_{n\ge 1}h^nB_n(f,g),
\end{equation}
where the $B_n$ are differential operators bilinear in $f$ and $g$, 
and $h$ is an expansion parameter.

Canonical case:
\begin{equation}
  f\ast g=fg+\sum_{n\ge 1}\frac{1}{n!}\left(\frac{i}{2}\right)^n\theta^{i_1j_1}\cdots\theta^{i_nj_n}(\pat_{i_1}\cdots\pat_{i_n}f)(\pat_{j_1}\cdots\pat_{j_n}g)
\end{equation}

Lie case:
\begin{eqnarray}
  f\ast g&=&fg+\sum_{n\ge 1}\frac{1}{n!}\left.\left(\frac{i}{2}\sum_kx^kg_k(i\pat_y,i\pat_z)\right)^nf(y)g(z)\right|_{y=x\atop z=x}\nn
  &=&fg+\frac{i}{2}x^kC^{ij}{}_k\pat_if\pat_jg+\ldots
\end{eqnarray}

Quantum space case ($h=\ln q$):
\begin{equation}
%  f\diamond g
%=fg+\sum_{n\ge 1}\frac{1}{n!}\left.\left(-hy\frac{\pat}{\pat y}x'\frac{\pat}{\pat x'}\right)^nf(x,y)g(x',y')\right|_{y'=y \atop x'=x}
f\diamond g
=fg+\sum_{n\ge 1}\frac{1}{n!}
(-h)^n\Big(\left(y\pat_y\right)^n f\Big)\Big(\left(x\pat_x\right)^n g\Big).
\end{equation}
The identification with formula (\ref{def_qua}) is obvious. 
In the following we shall work to second order in $h$ only. 
For the canonical and the Lie structure the formula for the 
$\ast$ commutator is
\begin{equation}
  \label{starcom}
  [f\ds g]=i\theta^{ij}(x)\pat_if\pat_jg+{\cal O}(\theta^3).
\end{equation}
This expression does not contain any terms in second order in $\theta$. 
This is typical for a deformation quanization of a Poisson structure \cite{Kontsevich}.

As a consequence the second term on the righthand side of (\ref{dia_delta_A}) 
will be:
\begin{equation}
  \label{x_starcom}
  [x^i\ds\al]=i\theta^{ij}\pat_j\al.
\end{equation}
For the canonical and the Lie structure (\ref{x_starcom}) will be true 
to all orders in $\theta$.
Combining (\ref{x_starcom}) and (\ref{starcom}) we obtain for (\ref{dia_delta_A})
\begin{equation}
  \delta A^i=\theta^{ij}\pat_j \alpha-\theta^{ij}\pat_i f \pat_j g+{\cal O}(\theta^3).
\end{equation}

Following Seiberg and Witten we construct explicitely  local expressions $A$ and 
$\alpha$ in terms of $a$, $\ve$ and $\theta$. This we do by the following Ansatz:
\begin{eqnarray}
  \label{poissonansatz}
  A^i&=&\theta^{ij}a_j+G^i(\theta,a,\pat a,\ldots)+{\cal O}(\theta^3)\\
  \alpha&=&\ve+\gamma(\theta,\ve,\pat\ve,\ldots,a,\pat a,\ldots)+{\cal O}(\theta^2).\nonumber
\end{eqnarray}
We demand that the variaton $\delta A$ of (\ref{poissonansatz}) with the 
infinitesimal parameter $\alpha$ is obtained from the variation  (\ref{ordg}) of a. 
This is true to first order in $\theta$ due to the Ansatz (\ref{poissonansatz}). 
In second order we get an equation for $G^i$ and $\gamma$:
\begin{eqnarray}
  \label{lie_beding}
  \delta_{\ve} G^i&=&\theta^{ij}\pat_j\gamma -\frac{1}{2}\theta^{kl}\left(\pat_k\ve\pat_l(\theta^{ij}a_j)+\pat_k(\theta^{ij}a_j)\pat_l\ve\right)\nn
  &&+i[\ve,G^i]+i[\gamma,\theta^{ij}a_j].
\end{eqnarray}
This equation has the following solution:
\begin{eqnarray}
  G^i&=&-\frac{1}{4}\theta^{kl}\{a_k,\pat_l(\theta^{ij}a_j)+\theta^{ij}F_{lj}\}\\
  \gamma&=&\frac{1}{4}\theta^{lm}\{\pat_l\alpha, a_m\},\nonumber
\end{eqnarray}
where $F_{ij}$ is the classical field strength $F_{ij}=\pat_ia_j-\pat_ja_i+i[a_i,a_j]$. 
To proof, that this indeed solves eqn (\ref{lie_beding}), one has to use 
the Jacobi identity for $\theta^{ij}(x)$. In the canonical case, i.e. 
$\theta^{ij}$ constant, this is the same result as found in \cite{SW}, 
if one takes into account the identification (\ref{hata}).  

Our quantum space example does not fit into the fromework of
deformaiton quantization as specified by eqn (\ref{starcom}), a 
quadratic term in $h=\ln q$ appears:
\begin{eqnarray}
  [f\dc g]&=&hxy(\pat_xf\pat_yg-\pat_xg\pat_yf)\\
  &&+\frac{h^2}{2}xy\left\{(\pat_yf\pat_xg-\pat_yg\pat_xf)+xy(\pat^2_yf\pat^2_xg-\pat^2_yg\pat^2_xf)\right.\nn
      &&\qquad\qquad\left.+x(\pat_yf\pat^2_xg-\pat_yg\pat^2_xf)+y(\pat^2_yf\pat_xg-\pat^2_yg\pat_xf)\right\}\nonumber
\end{eqnarray}
This has as a consequence that a second order term will appear 
in the following formula:
\begin{eqnarray}
  [x\dc\alpha]&=&+hxy\pat_y\alpha-\frac{h^2}{2}xy\pat_y(y\pat_y\alpha)\\{}  
  [y\dc\alpha]&=&-hxy\pat_x\alpha+\frac{h^2}{2}xy\pat_x(x\pat_x\alpha).\nonumber
\end{eqnarray}
Nevertheless the Seiberg-Witten map can be constructed at least for the 
abelean case. The transformation is
\begin{eqnarray}
  A^x&=& -ihxya^y-\frac{1}{2}h^2xy\left[\pat_y(xa^x(i-ya^y))+\pat_x(xya^ya^y)\right]+{\cal O}(h^3)\nn
  A^y&=& +ihxya^x-\frac{1}{2}h^2xy\left[\pat_x(ya^y(i-xa^x))+\pat_y(xya^xa^x)\right]+{\cal O}(h^3)\nn
  \alpha&=&\ve+\frac{1}{2}h\left[y\pat_y\alpha+x\pat_x\alpha+ixy(a_x\pat_y\alpha-a_y\pat_x\alpha)\right]+{\cal O}(h^2).
\end{eqnarray}

This sugggests that there should be an underlying geometric reason 
for the Seiberg-Witten map.

\begingroup\raggedright\endgroup

\end{document}